\documentclass[journal]{IEEEtran}
\usepackage{amsfonts}
\usepackage{amssymb}
\usepackage{graphicx}

\newtheorem{theorem}{Theorem}
\newtheorem{definition}{Definition}

\newtheorem{proposition}{Proposition}

\ifCLASSINFOpdf
\else
\fi
\usepackage[cmex10]{amsmath}
\begin{document}
%
\title{Space-Time Codes from Spectral Norm: A Fresh Look}
%
%
%

\author{Carlos A. R. Martins and
        Eduardo Brandani da Silva 
\thanks{C. A. R. Martins is with the Department
of Mathematics, UTFPR, Pato Branco, PR, Brazil e-mail: carlos@utfpr.edu.br.}
\thanks{E. B. Silva is with Department of Mathematics, Maringa State University, Maringa, PR 87020-900, Brazil e-mail: ebsilva@wnet.com.br.}
}

%
%

\markboth{Journal of \LaTeX\ Class Files,~Vol.~11, No.~4, December~2012}%
{Shell \MakeLowercase{\textit{et al.}}: Bare Demo of IEEEtran.cls for Journals}
%



\maketitle

\begin{abstract}
Current research proposes a natural environment for the space-time codes and in this context it is obtained a new design criterion for space-time codes in multi-antenna communication channels. The objective of this criterion is to minimize the pairwise error probability of the maximum likelihood decoder, endowed with the matrix spectrum norm. The random matrix theory is used and an approximation function for the probability density function is obtained for the largest eigenvalue of a Wishart Matrix.
\end{abstract}


%
\IEEEpeerreviewmaketitle

\section{Introduction}

During the last decade, we witnessed a huge expansion in the telecommunication area, but these are insufficient for the future. This is due to the growing need to develop reliable communication systems that allow high rates of data transmission. This has led to the study and development of new mathematical methods and structures that provide support for new technologies, mainly the wireless communication systems.

A consistent theory for communication systems was introduced by the classical work of Claude Elwood Shannon \cite{Sha48}, where it was proved that for any communication channel with capacity $C$, the data transmission below $C$ may be done efficiently using an appropriate error correcting code, that is, there exists a code such that the error probability can be as small as required. The results of Shannon's work are valid for channels with only one transmitter antenna and only one receiver antenna, called {\it SISO - Single Input and Single Output} channel.

Wireless transmissions occur under severe impairments, which restrict communication speed and reliability. {\it MIMO - Multiple Input and Multiple Output} communication systems with {\it  space=time codes} increase the reliability and the capacity of a transmission system without increasing the band. The works of Foschini-Gans \cite{FG98} and Telatar \cite{Tel99}  proved that the MIMO communication systems attained capacity of data transmission greater than SISO systems. These results proved the superiority of MIMO systems and called the attention of researchers, originating many results and real applications.

The foundations of space-time codes were established by Tarokh, Seshadri and Calderbank in \cite{TSC98} and space-time trellis codes (STTC) were introduced. Alamouti in \cite{Ala98} reduced the decoding complexity of the code presented in \cite{TSC98}, inventing a simple decoding process for two transmitter antennas. As in \cite{TSC98}, Alamouti obtained diversity in space and time in a very efficient way. Tarokh, Jafarkhani and Calderbank extended Alamouti's result to any number of transmitter antennas in \cite{TJC99}, inserting Alamouti's code in a new class of codes called space-time bloc codes (STBC). These research works gave origin to research about searching criteria to design good space-time codes for MIMO channels. In \cite{TSC98} and \cite{YCV03} we have two of the most important criteria used.

Current research proposes a new approach on STC and a new criterion for the design of STC. To obtain this criterion we introduced a new norm in the maximum likelihood decoder and we used several  results from the theory of Random Matrices.

The theory of random matrices emerged in the work of the mathematical statistician John Wishart in \cite{Wis28}, but it gained great visibility in the 1950s with the contributions of the physical mathematician Eugene Paul Wigner with the works \cite{Wig51}, \cite{Wig55} and \cite{Wig58}, about the spreading of resonance of particles with heavy cores in slow nuclear reactions. Further, the physical mathematician Freeman Dyson formalized the theory in \cite{Dys62a}, \cite{Dys62b} and \cite{Dys62c}. The theory of random matrices is used in several area and problems, as Riemann hypothesis, stochastic differential equation, statistic physics, chaotic systems, numerical linear algebra, neural networks, information theory, signal processing and the study of capacity of data transmission in MIMO channels. The deep mathematical results may be seen in the works \cite{TV10}, \cite{TV11a}, \cite{TV11b} and \cite{TV12}. For the applications see \cite{For10} and \cite{TVer04}.

The study of probability density and distribution functions of eigenvalues is one of the main problem in the random matrices theory and it was attacked in works by von Neumann, Birkhoff, Smale, Demmel and others. The pdf of the eigenvalues of a Wishart matrix was established in 1939 in \cite{Hsu39}. Many researchers studied this issue, for instance, \cite{Jon82}, \cite{MP67}, \cite{Tro84} and \cite{Wac78}. Estimations for the largest and smallest eigenvalues are given in \cite{Ger80}, \cite{Sug67}, \cite{KC71} and \cite{Sil85}. These results have many applications, and they were used to study channel capacity of MIMO channels, \cite{TVer04}, \cite{Tel99}, \cite{ALTV04} and \cite{CWZ03}.

The work is divided into five sections. Section two describes the fundamentals of STC and gives the design criteria to project STC. In this section we propose a natural environment where the STC living in and the maximum likelihood decoder is endowed with the spectral norm. In the third section we have the Random Matrices theory, with a focus on the cumulative distribution of eigenvalues of Wishart Matrices. Section four describes an approximation function to the probability density function for the largest eigenvalue of a Wishart Matrix, and this approximation will be used in the last section to obtain a new design criterion to project STC.

\section{Space-time codes and MIMO communication systems}

In this section we give a short review of space-time codes and MIMO channels. For more details see [32]. In short, these codes are a set of several techniques to achieve high rates of transmission. These rates are close to the theoretical limit of MIMO channels.

Let us consider a communication system with $n_T$ transmitter antennas and $n_R$ receiver antennas. In each time instant $t$, we have $n_T$ simultaneous transmissions, denoted by
\begin{eqnarray*}
c_t^1, c_t^2, \cdots, c_t^{n_T} \,,
\end{eqnarray*}
where $c_t^i$, a complex number, is the transmitted signal by the antenna $i$ in the time $t$, $1 \leq i \leq n_T$.

In each time instant $t$, the received signal by the receiver antenna $j$ is given by the following linear combination
\[
r_t^j = \sqrt{E_s} \sum_{i=1}^{n_T} h_{ji} (t) c_t^i + n_t^j, \hspace{1cm} j= 1, 2, \cdots, n_R \,.
\]
where $E_s$ is the average power by signal in each transmitter antenna and $n_t^j$ is the additive gaussian write noise for each receiver antenna $j$, where it is suppose that $n_t^j$ is a complex gaussian random variable with mean zero and  variance $N_0 /2$ by dimension. The coefficient $h_{ji} (t)$ is the fading between transmitter antenna $i$ and receiver antenna $j$. It is also supposed that $h_{ji} (t)$ are complex gaussian random variables with mean $m_{h_{ji} }$ and variance $1/2$ by dimension, which may be modeled as Rayleigh or Rice random variables.

The codewords, the coefficients $h_{ji} (t)$, the gaussian noises and the received signals, may be expressed by using matrices: $\textbf{C} =\left( c_{ij} \right)_{ij}$, $\textbf{H} =\left( h_{ki} \right)_{ki}$, $\textbf{N} =\left( n_{j}^{k} \right)_{kj}$ and $\textbf{R} =\left( r_{j}^{k} \right)_{kj}$, for $1 \leq i \leq n_T$, $1 \leq j \leq l$ and $1 \leq k \leq n_R$. Thus, we may write
\begin{equation}\label{mimo}
R=\sqrt{E_s} H C +N.
\end{equation}

In current work we are considering the coherent case, in which it is supposed that the fading $h_{ji} (t)$ is constant during $l$ time instants (constant in a frame), $h_{ji} (1) = h_{ji} (2) = \cdots =h_{ji} (l) = h_{ji}$.

Given the codeword ${\bf C}=(c_1^1 c_1^2 \cdots c_1^{n_T}   c_2^1 c_2^2 \cdots c_2^{n_T}  \cdots c_l^1 c_l^2 \cdots c_l^{n_T})$ was transmitted, where $l$ is the length of the frame, let us assume that the maximum likelihood decoder decides, wrongly, that the correct codeword sent was

${\bf E} = (e_1^1 e_1^2 \cdots e_1^{n_T}   e_2^1 e_2^2 \cdots e_2^{n_T} \cdots  e_l^1 e_l^2 \cdots e_l^{n_T})$. \\

\begin{definition}
The pairwise error probability, denoted by $P(\textbf{E} \rightarrow \textbf{C})$, is the probability of sending a codeword $C$ and it is decoded by $E$.
\end{definition}

Decoding by maximum likelihood minimizes the pairwise error probability. We suppose a total knowledge of the channel, coherent  channel or CSI, that is, the fading is known, and consequently, $H$. In this way, the maximum likelihood decoder may be interpreted as the probability of
\begin{equation}\label{desi01}
\parallel R-\sqrt{E_s} HE \parallel^2  \hspace{2mm} \leq \hspace{2mm}  \parallel R- \sqrt{E_s} HC \parallel^2
\end{equation}
occurs. We define the matrix $A(c,e) = \left( \sum_{t=1}^{l} (e_t^i -c_t^i)   \overline{(e_t^{k} -c_t^{k})} \right)_{i,k} ,  1 \leq i, k \leq n_T$ with rank $r$ and not null eigenvalues $\lambda_1, \lambda_2, \cdots, \lambda_r $. From \cite{TSC98}, one has
\begin{equation}\label{desi9}
P(C \rightarrow E \mid H)
\leq   \displaystyle \frac{1}{2} \prod_{j=1}^{n_R} \prod_{i=1}^{r} \exp \left(   -  \lambda_i \mid \beta_{ji} \mid^2   \frac{ E_s }{4 N_0} \right).
\end{equation}

One of the main results from \cite{TSC98} is a search criterion for STC. Today, this is known by {\bf Rank and Determinant Criterion}. The expression in (2.3) is hard to manipulate. Using some approximations, it may be written in a simpler way and the error probability is given by
\begin{equation}\label{criteposto1}
P(C \rightarrow E) \leq \left( \displaystyle \prod_{i=1}^{r}  \lambda_i  \right)^{-n_R} \left( \displaystyle \frac{E_s}{4 N_0} \right)^{-r n_R}.
\end{equation}

Therefore, the search of good space-time for wireless channels, when $r \cdot n_R$ is small ($\leq 4$), must be done to minimize (\ref{criteposto1}) and the criterion is given by:

\begin{center}
$\bullet$ To maximize the minimum rank $r$ of $A(c,e)$, on all pairs of distinct codewords;
\end{center}

\begin{center}
$\bullet$ To maximize the product $ \prod_{i=1}^{r}  \lambda_i$ of eigenvalues of $A(c,e)$, between all pairs of distinct codewords.
\end{center}

Another important search criterion for STC is also obtained from (\ref{desi9}), when $r \cdot n_R > 4$, established in \cite{YCV03}. Supposing the space-time code operates with reasonable $SNR$, after some approximations,
\begin{equation}\label{crittraco1}
P(C \rightarrow E) \leq \frac{1}{4}  \exp \left(  - \frac{E_s}{4 N_0} n_R  \sum_{i=1}^{r} \lambda_i \right).
\end{equation}

In this case, the searching of STC, when $r \cdot n_R$ is large ($\geq 4$), must minimize (\ref{crittraco1}). The limiting $(\ref{crittraco1})$ shows that the error probability is dominated by codewords with minimum sum of eigenvalues of $A(c,e)$, that is, $tr(A(c,e))$. Thus, the minimum sum of all eigenvalues of $A(c,e)$, between all pairs of distinct codewords must be maximized. This criterion is called {\bf Trace Criterion} and is given by:

\begin{center}
$\bullet$ The minimum rank $r$ of $A(c,e)$, over all pairs of distinct codewords satisfy that $r n_R \geq 4$.
\end{center}

\begin{center}
$\bullet$ To maximize the minimum trace $\sum_{i=1}^{r}  \lambda_i$ of $A(c,e)$ between all pairs of distinct codewords with minimum rank.
\end{center}

In connection with the two criteria given, it must be observed that the vast majority of works on STC deal with the search of new codes. In \cite{TSC98}, \cite{YCV03} and all other works, a vector and a matrix are seen as the same object, that is, a matrix is a representation of a vector from some space ${\bf R}^n$. However, from a mathematical point of view, there exist deep analytical, algebraic and geometric differences when a codeword $C$ is seen as a vector $\textbf{C}=(c_1^1 c_1^2 \cdots c_1^{n_T}   c_2^1 c_2^2 \cdots c_2^{n_T}  \cdots c_l^1 c_l^2 \cdots c_l^{n_T})$ or a matrix $\textbf{C} =\left( c_{ij} \right)_{ij}$. In equation (\ref{mimo}) it is fundamental to use multiplication of matrices, and there is not a similar multiplication property for vectors.

For the two criteria given, it is used freely both representations and the Frobenius norm is very useful, since $\parallel C \parallel_F^2$ of a matrix $C$ is the square of Euclidian norm of $C$, seen as a vector.

This approach is not the most appropriate.  If considered a convenient matrix space as a natural environment where space-time codes, gaussian noise and fading matrices living in, and if this matrix space has enough rich analytic, algebraic and geometric structures, we will have at hand powerful mathematical tools to manipulate the matrices. For instance, determinant, rank and trace, extensively used in space-time codes and MIMO works, are all operators on matrix spaces. The search for good STC is not done on vector spaces, neither in $\mathbb{C}^n$, as it is done in classic theory of error correcting codes, where $n_R = n_T=1$.

\begin{definition} Let $M(n,{\bf C})$ be the set of all $n \times n$ complex matrices. Under matrix addition and multiplication by complex numbers (scalars), $M(n,{\bf C})$ is a vector space. Together with matrix multiplication, it is a matrix algebra, that is, an associative algebra of matrices. The spectral norm on $M(n,{\bf C})$ is the function $\| \,.\,\|_2 : M(n,{\bf C}) \rightarrow [0,\infty)$, where given $A \in M(n,{\bf C})$ on has

\[
\|A\|_2 = \sqrt{\lambda_{max}(A^* A)} = \sigma_{max} (A) \,,
\]
where $A^*$, $\lambda_{max}(A)$ and $\sigma_{max}(A)$ are respectively, the conjugate transpose, the largest eigenvalue and the largest singular value of $A$. The spectral norm has the following fundamental properties for all matrices $A$ and $B$ in $M(n,{\bf C})$ and all scalar $\alpha$:
\begin{itemize}
\item[i)] $\|A\| \geq 0$

\item[ii)] $\|A\|=0 \iff A=0$

\item[iii)] $\|\alpha A\|=|\alpha |\|A\|$

\item[iv)] $\|A + B\| \leq \|A\| + \|B\|$ .

\item[v)] $\|A B\| \leq \|A\|\|B\|$
\end{itemize}
\end{definition}

The space $M(n,{\bf C})$ endowed with the spectral norm is a Banach algebra. \\

\begin{definition}
A {\bf space-time code} (STC) is a finite subset of $M(n,{\bf C})$.
\end{definition}

Definition (2.3) is very generic and to obtain applicable STC, subsets of $M(n,{\bf C})$ with geometric and algebraic properties must be considered. It is supposed that the average power by signal in each transmitter antenna is unitary, that is, $E_s = 1$. Thus, when the codeword $C$ is sent, the received signal is
\[
R=HC+N .
\]

If $E$ is wrongly chosen by the maximum likelihood decoder, when $E$ is received, one has
\begin{eqnarray*}
\parallel R-HE \parallel_2 & \leq & \parallel R-HC \parallel_2 \\
\parallel HC +N -HE \parallel_2 & \leq &\parallel HC +N-HC \parallel_2 \\
\parallel -H(E-C) +N  \parallel_2  & \leq & \parallel N \parallel_2 \\
\parallel H(E-C) -N  \parallel_2 & \leq & \parallel N \parallel_2 \\
\parallel H(E-C)  \parallel_2 - \parallel N  \parallel_2 & \leq & \parallel H(E-C) -N  \parallel_2 \hspace{1mm} \leq \hspace{1mm} \parallel N \parallel_2 ,
\end{eqnarray*}
and consequently
\[
\frac{1}{4} \parallel H(E-C)  \parallel_2^2 \hspace{2mm} \leq \hspace{1mm} \parallel N \parallel_2^2 \,.
\]

We want the error probability of the maximum likelihood decoder. Since $\| A \|_2^2 = \lambda_{max} (AA^{*})= \sigma_{max}^2 (A)$ and $H$ is known, we need to calculate
\begin{eqnarray*}
P(C \rightarrow E \mid H) &=& P(\parallel R-HE \parallel_2^2 \leq \parallel R-HC \parallel_2^2) \\
& \leq & P \left( \frac{1}{4} \parallel H(E-C)  \parallel_2^2 \leq \parallel N \parallel_2^2 \right) \\
& = & P \left( \frac{1}{4} \parallel H(E-C)  \parallel_2^2 \leq \lambda_{max} (NN^*)\right) \,.
\end{eqnarray*}
This implies that we need to find the pdf of the largest eigenvalue of $NN^{*}$. It is the subject of the next section.

\section{Random Matrices}

In his works, Wigner realized that the eigenvalues distribution of a matrix with random gaussian entries coincided with the statistics of fluctuations of the levels of heavy atoms, obtained experimentally. Thus, the pdf of eigenvalues of Random Matrices became an important object.

It is assumed that variance $\sigma^2 = 1$ in the definitions below, just for simplicity, but any variance could be used. The set of all random variables $z = x + iy$, where $x$ and $y$ are iid $N(\mu, \sigma^2)$, is denoted by $\tilde{N}(\mu, \sigma^2)$.

\begin{definition}
{\bf Complex Gaussian} $\tilde{G}(m,n)$ is the family of all $m \times n$ random matrices with independent and identically distributed (iid) elements which are $\tilde{N}(0,1)$.

{\bf Wishart} $\tilde{W}(m,n)$ is the family of all $m \times n$ symmetric random matrices, which may be written in the form $AA^{T}$, where  $A \in \tilde{G}(m,n)$.

{\bf Gaussian Unitary Ensemble} $GUE$ is the set of all symmetric $m \times m$ random matrices with (iid) elements that are  $N(0, 1/4)$ in the upper-triangle and iid elements that are $N(0, 1/2)$ on the diagonal.
\end{definition}

Usually the elements of a Wishart matrix are not iid. Then the joint distribution is more complicated, where Cholesky and L.U decompositions are used, see \cite{Mui82}.

Now, considering $\tilde{G}(m,n)$, where the elements of matrices are $\tilde{N}(0, \sigma^2)$, one has the following result from [8].

\begin{theorem}
Given $\tilde{M} = \tilde{A} \tilde{A^{*}} \in \tilde{W} (m,n)$, where $\tilde{A} \in \tilde{G}(m,n)$, suppose $\lambda_1 \geq \lambda_2  \geq \cdots \geq \lambda_{m-1} \geq \lambda_m \geq 0$ are the eigenvalues of $\tilde{M}$. Then, the joint pdf of the eigenvalues of $\tilde{M}$ is
\begin{equation}\label{fdpconjunrecom}
\begin{split}
\tilde{f}(\lambda_1, \lambda_2, \cdots, \lambda_m)=\tilde{K}_{n,m} \exp\left( -\frac{1}{2 \sigma^2} \sum_{i=1}^{m} \lambda_i \right) \times \\
\times \prod_{i=1}^{m} \lambda_i^{n-m} \prod_{i<j} (\lambda_i - \lambda_j)^2, \nonumber
\end{split}
\end{equation}
where
\begin{equation}\label{constanterecom}
\tilde{K}_{n,m}^{-1} =(2 \sigma^2)^{mn} \prod_{i=1}^{m} \Gamma \left( n-i+1 \right) \Gamma \left(m-i+1 \right).
\end{equation}
\end{theorem}

\section{An approximation to the largest eigenvalue pdf of a complex Wishart matrix}

To conclude section 2 we need the pdf of the largest eigenvalue of a complex Wishart matrix. The results found in literature, for instances, in \cite{ZC08} and \cite{ZCW09}, are not easy to manipulate. Thus, an approximation to the pdf is obtained, besides being convenient for our purposes, it has an independent interest. We begin with a bounding result similar to Lemma \ref{lema4.2Edel}, for $\tilde{W}(m,n)$.

\begin{theorem}
\label{new result}
If $\tilde{M}  \in \tilde{W}(m,n)$, then $f_{\lambda_{max}} (\lambda)$ satisfies
\begin{eqnarray*}
f_{\lambda_{max}} (\lambda) & \leq  &  \frac{\tilde{K}_{n,m}}{\tilde{K}_{n-1,m-1}}  \lambda^{n+m-2}  \exp\left( -\frac{\lambda}{2 \sigma^2} \right) \\
& = & \frac{(2 \sigma^2)^{-n-m+1}}{\Gamma(n) \Gamma(m)} \lambda^{n+m-2} \exp \left( - \frac{\lambda}{2 \sigma^2} \right).
\end{eqnarray*}
\end{theorem}

\begin{IEEEproof}
The joint pdf of eigenvalues of a Wishart matrix defined in (\ref{fdpconjunrecom}) is given by
\begin{equation}
\begin{split}
\tilde{K}_{n,m} \exp\left( -\frac{\lambda_1}{2 \sigma^2} \right) \lambda_{1}^{n-m} \exp\left(  \sum_{i=2}^{m} -\frac{\lambda_i}{2 \sigma^2} \right) \\
\times \prod_{i=2}^{m} \left( (\lambda_1 - \lambda_i)^2 \cdot \lambda_i^{n-m}  \right) \prod_{i<j} (\lambda_i - \lambda_j)^2 . \nonumber
\end{split}
\end{equation}
Thus,
\begin{eqnarray*}
f_{\lambda_{max}} (\lambda) & = & \tilde{K}_{n,m} \exp\left( -\frac{\lambda}{2 \sigma^2} \right) \lambda^{n-m} \int_{R}\exp\left(  \sum_{i=2}^{m} -\frac{\lambda_i}{2 \sigma^2} \right)  \\
& & \times \prod_{i=2}^{m}  \left( (\lambda - \lambda_i)^2 \cdot \lambda_i^{n-m} \right) \prod_{i<j} (\lambda_i - \lambda_j)^2 d\lambda_i,
\end{eqnarray*}
where $R=\{ (\lambda_2,  \lambda_3, \cdots , \lambda_{m}): \lambda_2 \in [0,\lambda] ; \lambda_i \in [0, \lambda_{i-1}], i \in \{3, \cdots,m \}\}$. Since $0 \leq \lambda -\lambda_i \leq \lambda$, then $\leq (\lambda -\lambda_i)^2 \leq \lambda^2$ and this may be bounded above by
\begin{eqnarray*}
f_{\lambda_{max}} (\lambda)  & \leq &  \tilde{K}_{n,m} \exp\left( -\frac{\lambda}{2 \sigma^2} \right) \lambda^{n-m} \int_{R}\exp\left(  \sum_{i=2}^{m} -\frac{\lambda_i}{2 \sigma^2} \right) \\
& & \times \prod_{i=2}^{m}  \left( \lambda^2 \cdot \lambda_i^{n-m} \right) \prod_{i<j} (\lambda_i - \lambda_j)^2 d\lambda_i \\
& \leq & \tilde{K}_{n,m} \exp\left( -\frac{\lambda}{2 \sigma^2} \right) \lambda^{n+m-2} \\
& & \times \int_{R_2}\exp\left(  \sum_{i=2}^{m} -\frac{\lambda_i}{2 \sigma^2} \right) \\
 & & \times \prod_{i=2}^{m}  \left( \lambda_i^{n-m} \right) \prod_{i<j} (\lambda_i - \lambda_j)^2 d\lambda_i,
\end{eqnarray*}
where $R_2=\{(\lambda_2, \cdots, \lambda_m): \lambda_2 \in [0,\infty]; \lambda_i \in [0,\lambda_{i-1}], i \in \{3,	 \cdots,m\}\}$. From Theorem 3.7, we have
\begin{equation}
\tilde{K}_{n-1,m-1}^{-1} = \int_{R_2} \exp\left( \sum_{i=2}^{m} \frac{-\lambda_i}{2 \sigma^2} \right) \prod_{i=2}^{m} \lambda_i^{n-m} \prod_{i<j} (\lambda_i - \lambda_j)^2 d\lambda_i \,,
\end{equation}

and substituting (4.1) in the limiting of $f_{\lambda_{max}} (\lambda)$ one has
\[
f_{\lambda_{max}} (\lambda) \leq \frac{\tilde{K}_{n,m}}{\tilde{K}_{n-1,m-1}} \lambda^{n+m-2}  \exp\left( -\frac{\lambda}{2 \sigma^2} \right).
\]
Finally, using the expression of $\tilde{K}_{n,m}$ in equation (\ref{constanterecom}), one has
\begin{eqnarray*}
f_{\lambda_{max}} (\lambda) & \leq &  \frac{(2 \sigma^2)^{-n-m+1}}{\Gamma(n) \Gamma(m)} \lambda^{n+m-2} \exp\left( -\frac{\lambda}{2 \sigma^2} \right).
\end{eqnarray*}
\end{IEEEproof}

\begin{figure}
\centering
\includegraphics[
height=2.0611in,
width=2.9331in
]{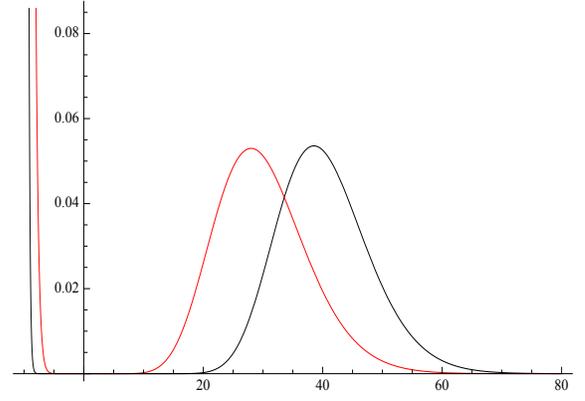}
\caption{Graphs of exact pdf $f_{\lambda_{max}}(\lambda)$ and of $g(\lambda)$ for $\tilde{W}(3,13)$}
\label{W313}
\end{figure}

The limiting of Theorem \ref{new result} is not a pdf, since
\small
\[
\int_{0}^{\infty} \frac{(2 \sigma^2)^{-n-m+1}}{\Gamma(n) \Gamma(m)} \lambda^{n+m-2} \exp\left( -\frac{\lambda}{2 \sigma^2} \right) d\lambda = \frac{\Gamma (n+m-1)}{\Gamma(n) \Gamma(m)} \neq 1 \,.
\]
\normalsize
Normalizing the limiting of Theorem \ref{new result}, we define the function
\begin{eqnarray}
g(\lambda) = \frac{(2 \sigma^2)^{-n-m+1}}{\Gamma(n+m-1)} \lambda^{n+m-2} \exp\left( -\frac{\lambda}{2 \sigma^2} \right) \,.
\end{eqnarray}
Then, $g$ is a pdf on $[0,\infty)$. Using an algebraic computer program, $f_{\lambda_{max}} (\lambda)$ may be plotted for all cases of $m$ and $n$. Comparing equivalent cases of $g$ with $f_{\lambda_{max}}$, it may be seen that a translation of $g(\lambda)$ is a good approximation to $f_{\lambda_{max}}(\lambda)$. In figure \ref{W313} we have an example.

\begin{figure}
\begin{center}
\includegraphics[
height=2.0611in,
width=2.9331in
]{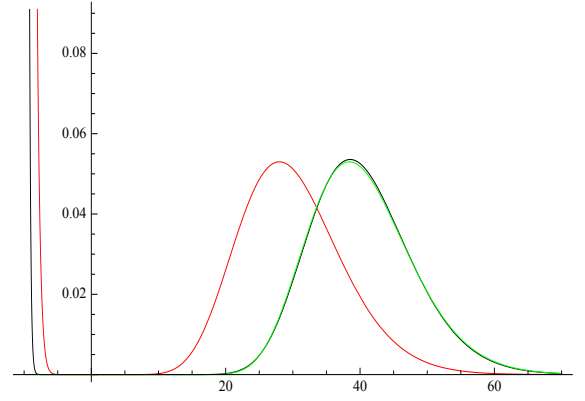}
\caption{Graphs of $f_{\lambda_{max}}(\lambda)$, $g(\lambda)$ and the translation in $10.4$ for $\tilde{W}(3,13)$}
\label{W313tr}
\end{center}
\end{figure}

Then, a constant $d_1 = d_1(m,n)$ must be found such that translation of $g(\lambda)$ is given by
\begin{eqnarray*}
 \phi(\lambda) =\left\{ \begin{array}{ccc}
\hspace{3cm} 0 \hspace{5cm} , \hspace{1cm} 0 \leq \lambda < d_1 \\
\displaystyle \frac{(2 \sigma^2)^{-n -m +1} \cdot \left( \lambda -d_1 \right)^{n + m -2}}{ \Gamma(n + m-1)}  \exp\left(-\frac{(\lambda -d_1)}{2 \sigma^2}\right)\hspace{1mm} , \hspace{1cm}  \lambda \geq d_1
\end {array}\right.,
\end{eqnarray*}
be an approximation to $f_{\lambda_{max}}(\lambda)$. The figure \ref{W313tr} shows the graphs of $f_{\lambda_{max}}(\lambda)$, $g(\lambda)$ and $\phi(\lambda -10.4)$ for $\tilde{W}(3,13)$.

Table \ref{tabela01} shows the exact pdf $f_{\lambda_{max}}(\lambda)$ for some cases and the translations of $g(\lambda)$ which better fit, such that $\phi (\lambda)$ is the best approximation. Data were obtained by trial and error to minimize the distance
\[
\int_{0}^{\infty} \mid f_1 (t) - f_2(t) \mid dt,
\]
between $f_{\lambda_{max}}(\lambda)$ and $\phi (\lambda)$. Table \ref{tabela01} also presents the maximum of $\phi(t)$. For simplicity, in Tables \ref{tabela01} and \ref{tabela02} it is assumed $\sigma^2 = 1$. Since $\phi(\lambda)$ is a translation of $g(\lambda)$, they have the same maximum.

\begin{table}
\centering
\begin{tabular}{|c|c|c|}
\hline
\rule[-1ex]{0pt}{3.5ex} $\tilde{W}(m,n)$ & better translation & maximum point of the better translation \\
\hline
\rule[-1ex]{0pt}{3.5ex} $\tilde{W}(2,2)$ & $1.0$ & $5.0$\\
\hline
\rule[-1ex]{0pt}{3.5ex} $\tilde{W}(2,3)$ & $1.7$ & $7.7$ \\
\hline
\rule[-1ex]{0pt}{3.5ex} $\tilde{W}(2,4)$ & $2.4$ & $10.4$ \\
\hline
\rule[-1ex]{0pt}{3.5ex} $\tilde{W}(2,15)$ & $6.7$ & $36.7$ \\
\hline
\rule[-1ex]{0pt}{3.5ex} $\tilde{W}(2,22)$ & $8.7$ & $52.7$ \\
\hline
\rule[-1ex]{0pt}{3.5ex} $\tilde{W}(3,3)$ & $3.2$ & $11.2$ \\
\hline
\rule[-1ex]{0pt}{3.5ex} $\tilde{W}(3,5)$ & $5.1$ & $17.1$ \\
\hline
\rule[-1ex]{0pt}{3.5ex} $\tilde{W}(3,13)$ & $10.4$ & $38.4$ \\
\hline
\rule[-1ex]{0pt}{3.5ex} $\tilde{W}(3,25)$ & $16.0$ & $68$ \\
\hline
\rule[-1ex]{0pt}{3.5ex} $\tilde{W}(4,4)$ & $5.5$ & $17.5$ \\
\hline
\rule[-1ex]{0pt}{3.5ex} $\tilde{W}(4,7)$ & $9.0$ & $27$ \\
\hline
\rule[-1ex]{0pt}{3.5ex} $\tilde{W}(4,13)$ & $14.0$ & $44$ \\
\hline
\rule[-1ex]{0pt}{3.5ex} $\tilde{W}(5,5)$ & $8.4$ & $24.4$ \\
\hline
\rule[-1ex]{0pt}{3.5ex} $\tilde{W}(5,9)$ & $13.2$ & $37.2$ \\
\hline
\end{tabular}
\caption{Better translation and its maximum point.}
\label{tabela01}
\end{table}

\begin{table*}
\centering
\begin{tabular}{|c|c|c|c|}
\hline
\rule[-1ex]{0pt}{3.5ex} $\tilde{W}(m,n)$ & maximum point $f_{\lambda_{max}}(\lambda)$ & maximum point of $g(\lambda)$ & difference between the maximum points\\
\hline
\rule[-1ex]{0pt}{3.5ex} $\tilde{W}(2,2)$ & $5.11$ & $4.0$ & $1.1$\\
\hline
\rule[-1ex]{0pt}{3.5ex} $\tilde{W}(2,3)$ & $7.89$ & $6.0$ & $1.89$\\
\hline
\rule[-1ex]{0pt}{3.5ex} $\tilde{W}(2,4)$ & $10.53$ & $8.0$ & $2.53$\\
\hline
\rule[-1ex]{0pt}{3.5ex} $\tilde{W}(2,15)$ & $36.7$ & $30.0$ & $6.7$ \\
\hline
\rule[-1ex]{0pt}{3.5ex} $\tilde{W}(2,22)$ & $52.58$ & $44.0$ & $12.58$ \\
\hline
\rule[-1ex]{0pt}{3.5ex} $\tilde{W}(3,3)$ & $11.22$  & $8.0$ & $3.22$ \\
\hline
\rule[-1ex]{0pt}{3.5ex} $\tilde{W}(3,5)$ & $17.24$ & $12.0$ & $5.24$ \\
\hline
\rule[-1ex]{0pt}{3.5ex} $\tilde{W}(3,13)$ & $38.52$ & $28.0$ & $10.52$\\
\hline
\rule[-1ex]{0pt}{3.5ex} $\tilde{W}(3,25)$ & $67.79$ & $52.0$ & $15.79$\\
\hline
\rule[-1ex]{0pt}{3.5ex} $\tilde{W}(4,4)$ & $17.76$ & $12.0$ & $5.76$\\
\hline
\rule[-1ex]{0pt}{3.5ex} $\tilde{W}(4,7)$ & $27.15$ & $18.0$ & $9.15$\\
\hline
\rule[-1ex]{0pt}{3.5ex} $\tilde{W}(4,13)$ & $44.06$  & $30.0$ & $14.06$\\
\hline
\rule[-1ex]{0pt}{3.5ex} $\tilde{W}(5,5)$ & $24.53$ & $16.0$ & $8.53$\\
\hline
\rule[-1ex]{0pt}{3.5ex} $\tilde{W}(5,9)$ & $37.6$ & $24.0$ & $13.6$\\
\hline
\end{tabular}
\caption{Maximum points and their differences.}
\label{tabela02}
\end{table*}

It is not always possible to find the maximum point of $f_{\lambda_{max}}(\lambda)$. However, the maximum point of $g(\lambda)$ may be found. If this way, given the maximum of $g(\lambda)$, we must determine the constant $d_1 = d_1(m,n)$. Supposing $\sigma^2 =1$, the maximum point of $g(\lambda)$ is $\lambda_0 = 2 (n+m-2)$. Thus, $\lambda_0 + d_1(m,n) = 2 (n+m-2) + d_1(m,n)$ must coincide with the maximum point of $f_{\lambda_{max}}(\lambda)$. Let $h(m,n)$ be the maximum point of $f_{\lambda_{max}}(\lambda)$, then

\[h(m,n)=2 (n+m-2) + d_1(m,n)
\]
and
\[
d_1(m,n)=h(m,n) -2 (n+m-2) .
\]
Using the data of Table \ref{tabela02}, an expression to $d_1(m,n)$ will be obtained using the least squares method. Plotting the data of Table \ref{tabela02}, the function describing $d_1(m,n)$ may be seen as a plane and its equation given by
\[
d_1(m,n) = am+bn+c \,,
\]
where
\[
d_1(m_i,n_i)= \mu_i,
\]
and $\mu_i$ are the data of the third column of Table \ref{tabela02}. Thus, we must find the constants $a, b$ and $c$ by minimizing:
\[
F(a,b,c)= \sum_{i=1}^{14} (am_i+bn_i+c -\mu_i)^2.
\]
Then, we need to solve the equation $\nabla F(a,b,c)=0$, given by
\begin{eqnarray*}
\left\{ \begin{array}{ccc}
2 \displaystyle \sum_{i=1}^{14} (am_i+bn_i +c-\mu_i)\cdot (m_i)=0
\\
2 \displaystyle \sum_{i=1}^{14} (am_i+bn_i +c-\mu_i)\cdot (n_i)=0
\\
2 \displaystyle \sum_{i=1}^{14} (am_i+bn_i +c-\mu_i)\cdot (1)=0
\end {array}\right. .
\end{eqnarray*}
From the Table \ref{tabela02}, one has
\begin{eqnarray*}
 \left\{ \begin{array}{ccc}
154 \cdot a + 396 \cdot b + 44 \cdot c=  380.44
\\
396 \cdot a + 1906 \cdot b + 130 \cdot c=  1397.54
\\
44 \cdot a + 130 \cdot b + 14 \cdot c=  110.67
\end {array}\right.,
\end{eqnarray*}
and the solution is given by
\[
\{a,b,c \}=\{2.53573 \hspace{1mm},\hspace{1mm} 0.574893 \hspace{1mm},\hspace{1mm} -5.40273 \}.
\]

Therefore, the translation of $g(\lambda)$ is
\begin{eqnarray}\label{d1mm}
d_1 (m,n)=2.53573 m + 0.574893 n -5.40273.
\end{eqnarray}

{\bf Remark}. Table \ref{tabela03} shows values of \ref{d1mm}, and may be compared with the values of Tables \ref{tabela01} and \ref{tabela02}. We have
\[
\bar{\mu}=\frac{1}{14} \sum_{i=1}^{14} \mu_i =7.905 \,.
\]
Then, the \textit{total variation} is
\[
\sum_{i=1}^{14} (\mu_i -\bar{\mu})^2 =309.68,
\]
and the \textit{explained variation} is
\[
\sum_{i=1}^{14} (d_1(m_i,n_i) -\bar{\mu})^2 =290.364.
\]
Therefore, the \textit{coefficient of determination} is $R^2 = 296.364/309.68=0.953994$, implying that the model explains the observed values with $95\%$ of confidence.

Putting together the results, one has
\begin{table}
\centering
\begin{tabular}{|c|c|}
\hline
\rule[-1ex]{0pt}{3.5ex} $ d_1 (m,n)=2.53573 m + 0.574893 n -5.40273$ & valor num\'erico \\
\hline
\rule[-1ex]{0pt}{3.5ex} $d_1(2,2)$ & $0.8185$ \\
\hline
\rule[-1ex]{0pt}{3.5ex} $d_1(2,3)$ & $1.3934$ \\
\hline
\rule[-1ex]{0pt}{3.5ex} $d_1(2,4)$ & $1.9683$ \\
\hline
\rule[-1ex]{0pt}{3.5ex} $d_1(2,15)$ & $8.2921$ \\
\hline
\rule[-1ex]{0pt}{3.5ex} $d_1(2,22)$ & $12.3164$ \\
\hline
\rule[-1ex]{0pt}{3.5ex} $d_1(3,3)$ & $3.9291$ \\
\hline
\rule[-1ex]{0pt}{3.5ex} $d_1(3,5)$ & $5.0789$ \\
\hline
\rule[-1ex]{0pt}{3.5ex} $d_1(3,13)$ & $9.6780$ \\
\hline
\rule[-1ex]{0pt}{3.5ex} $d_1(3,25)$ & $16.5768$ \\
\hline
\rule[-1ex]{0pt}{3.5ex} $d_1(4,4)$ & $7.0397$ \\
\hline
\rule[-1ex]{0pt}{3.5ex} $d_1(4,7)$ & $8.7644$ \\
\hline
\rule[-1ex]{0pt}{3.5ex} $d_1(4,13)$ & $12.2138$ \\
\hline
\rule[-1ex]{0pt}{3.5ex} $d_1(5,5)$ & $10.1504$ \\
\hline
\rule[-1ex]{0pt}{3.5ex} $d_1(5,9)$ & $12.45$ \\
\hline
\end{tabular}
\caption{Values for the translation.}
\label{tabela03}
\end{table}

\begin{theorem}
\label{fdpNN*}
An approximation to the pdf of the largest eigenvalue of a Wishart matrix $N_{n_R \times l}N_{l \times n_R}^*$, with variance $\sigma^2 =N_0 /2$, is the given by the function
\small
\begin{equation}
\phi(t)  = \left\{ \begin{array}{ll}
0 &  , 0 \leq t< d_1 \\
\displaystyle \left( \frac{ t-d_1}{N_0}\right)^{ l+n_R -2} \frac{1}{\Gamma(l+n_R -1) \cdot N_0} \\
\times \exp\left(-\frac{(t-d_1)}{N_0}\right) & , t \geq d1
\end {array}\right. , \nonumber
\end{equation}
\normalsize
where $d_1=d_1 (n_R,l)=2.53573 n_R + 0.574893 l -5.40273$.
\end{theorem}

\section{A New Criterion to search STC}

The use of random matrices to obtain a search criteria of STC for MIMO channels is unknown to us. The results from Section 4 will establish such a one. From Section 2 we need to calculate
\begin{eqnarray}\label{probpesqui}
P \left(  \frac{1}{4} \parallel H(E-C) \parallel_2^2 \hspace{1mm} \leq \hspace{1mm} \lambda_{max} (NN^{*}) \right) \,,
\end{eqnarray}
and
\begin{eqnarray*}
P(a \leq \lambda_{max} (NN^{*}) )= \int_{a}^{\infty} f_{\lambda_{max}} (\lambda) d \lambda,
\end{eqnarray*}
where $f_{\lambda_{max}} (\lambda)$ is the pdf of the largest eigenvalue of a Wishart matrix. From theorem \ref{fdpNN*}, we consider that
\begin{eqnarray*}
P(a \leq \lambda_{max} (NN^{*}) )= \int_{a}^{\infty} \phi (t) dt .
\end{eqnarray*}

If $0 \leq a \leq d_1$, then
\begin{eqnarray*}
\int_{a} ^{\infty} \phi(t) dt & = 1.
\end{eqnarray*}
Equation (\ref{probpesqui}) is the probability of the maximum likelihood decoder, when receiving $R$, choose wrongly $E$, if $C$ was sent. When $0 \leq a < d_1$, an error occurred. On the other hand, if $ a \geq d_1$,
\begin{eqnarray*}
\displaystyle \int_{a} ^{\infty} \phi(t) dt & = & \displaystyle \int_{a} ^{\infty}  \frac{(N_0)^{-l-n_R +1} \cdot  \left( t-d_1 \right)^{(l+n_R -2)}}{\Gamma(l+n_R -1)} \\
& & \times \exp\left(-\frac{t-d_1}{N_0}\right) dt \\
& = & \displaystyle \frac{1}{\Gamma(l+n_R -1)}  \Gamma \left( l+n_R -1, \frac{a-d_1}{N_0} \right).
\end{eqnarray*}

Hence,
\small
\begin{equation}
P \left( a \leq  \lambda_{max} (NN^*)\right) = \left\{ \begin{array}{ll}
1 & , 0 \leq a \leq  d_1 \\
\displaystyle \frac{\Gamma \left(  l+n_r -1, \frac{ a -d_1}{N_0}\right)}{\Gamma \left(  l+n_r -1\right)} \hspace{3mm} & , d_1 < a
\end{array}\right., \nonumber
\end{equation}
\normalsize
where $d_1=d_1(m,n)$ is given by (\ref{d1mm}). Therefore, we proved the following:

\begin{theorem}
\label{teofaltatirarmediaH}
In a MIMO communication channel, given that the codeword $C$ was sent, if the maximum likelihood decoder is endowed with the spectral norm, the error probability of received signal be decoded by the codeword $E$, given that $H$ is known, is
\small
\begin{eqnarray*}
\lefteqn{P(C \rightarrow E \mid H)  = P \left( \frac{1}{4} \parallel H(E-C)  \parallel_2^2 \hspace{2mm} \leq \hspace{1mm}  \lambda_{max} (NN^*)\right)} & &   \\
& = & \left\{ \begin{array}{ll}
1 & , 0 \leq \frac{\parallel H(E-C)  \parallel_2^2}{4}  \leq  d_1 \\ \displaystyle \frac{\Gamma \left(  l+n_r -1, \frac{ \parallel H(E-C)  \parallel_2^2 -4 d_1}{4 N_0}\right)}{\Gamma \left(  l+n_r -1\right)} & , d_1 < \frac{\parallel H(E-C)  \parallel_2^2}{4}
\end {array}\right. ,
\end{eqnarray*}
\normalsize
where $d_1=d_1(n_R,l)=2.53573 n_r + 0.574893 l -5.40273$.
\end{theorem}

Until now, it is supposed that $H$ is known, that is, the statistics of $H$ are known. Now, we want to calculate the mean in $H$, that is,
\begin{eqnarray}\label{media}
P(C \rightarrow E )= \displaystyle \int_{Dom \hspace{0.3mm} p(H)} P(C \rightarrow E \mid H) p(H) dH,
\end{eqnarray}
where $p(H)$ is a pdf of the matrix $H$.

From Theorem 5.1, the term $\|H(E-C)\|_2$ is our main concern, since we need more information on the term $\Gamma \left(l+n_r -1, \frac{ \parallel H(E-C)  \parallel_2^2 -d_1}{4 N_0}\right)/\Gamma \left(l+n_r -1\right)$. From property $v$ of spectral norm, one has
\[
\displaystyle \frac{\parallel H(E-C)  \parallel_2^2}{4 N_0} -\frac{d_1}{N0} \leq \frac{\parallel H \parallel_2^2 \cdot \parallel (E-C) \parallel_2^2}{4 N_0} -\frac{d_1}{N_0}.
\]
Let $t = \parallel H \parallel_2^2 $ and $c = \parallel (E-C) \parallel_2^2$, then
\[
\frac{\parallel H \parallel_2^2 \cdot \parallel (E-C) \parallel_2^2}{4 N_0} -\frac{d_1}{N_0} = \frac{ t \cdot c}{4 N_0} -\frac{d_1}{N_0}= \frac{t \cdot c - 4 \cdot d_1}{4 N_0}.
\]

Now, define $f_m(x) = \Gamma(m,x)/\Gamma(m)$ for $m > 0$ fixed and $x \geq 0$. A typical example is shown in Figure 3. We know that $f_m(x)$ is a fast decreasing function, such that $0 < f_m(x) \leq 1$, $\lim_{x \rightarrow \infty} f_m(x) = 0$ and $\lim_{m  \rightarrow \infty} f_m(x) = 1$, for all $x$.

Our goal is to achieve a criteria to search STC, thus from the behavior of $f_m(x)$ it will be enough assume the following
\begin{eqnarray}
\displaystyle P(C \rightarrow E \mid H) & \approx & \left\{ \begin{array}{ll}
1 & , 0 \leq t \leq \frac{4 \cdot d_1}{c}
\\
\frac{\Gamma \left(  l+n_r -1, \frac{t \cdot c - 4 \cdot d_1}{4 N0} \right)}{\Gamma \left(  l+n_r -1\right)} \hspace{3mm} & , \frac{4 \cdot d_1}{c} < t < \infty
\end {array}\right..
\label{teofaltatirarmediaHaproxi}
\end{eqnarray}

\begin{figure}
\begin{center}
\includegraphics[height=2.0611in,
width=2.9331in
]{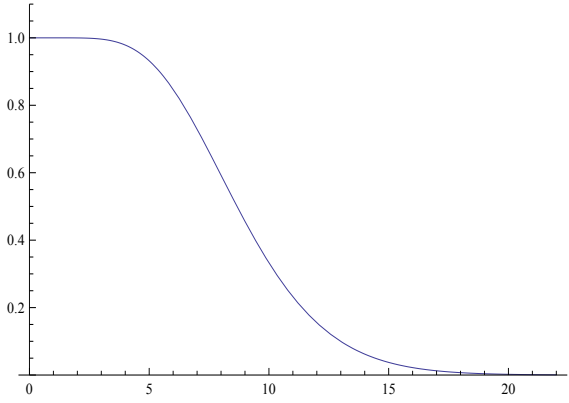}
\caption{graph of $f_m(x)$ for $m = 9$}
\label{examplef}
\end{center}
\end{figure}

The elements of $H_{n_R \times n_T}$ are gaussian random complex variables with mean zero and variance $1/2$. From Theorem 4.2, the pdf of the largest eigenvalue of $H_{n_R \times n_T}H_{n_T \times n_R}^*$ is given by
\begin{equation}
\label{fdpHH*}
\psi(t) =\left\{\begin{array}{ll}
0 & , t < d_2 \\
\frac{\left( t-d_2 \right)^{n_T+n_R -2}}{ \Gamma(n_T+n_R -1)}  \exp\left(-t+d_2 \right) & , d_2 \leq t
\end {array}\right. ,
\end{equation}

where $d_2 =d_2(n_R, n_T)= 2.53573 n_R + 0.574893 n_T -5.402373$. With equations (\ref{teofaltatirarmediaHaproxi}) and (\ref{fdpHH*}), the probability $P(C \rightarrow E )= \displaystyle \int_{Dom \hspace{0.3mm} \psi(t)} P(C \rightarrow E \mid H) \psi(t) dt$, is given by
\small
\begin{equation} \label{integralfinal}
\int_{d_2}^{\infty}  \frac{\Gamma(l+n_R-1, \frac{ c \cdot t -4 d_1}{4 N_0})}{\Gamma(l+n_R-1)}
\left[ \frac{\left( t-d_2 \right)^{n_T+n_R -2} \exp\left(-t+d_2\right)}{ \Gamma(n_T+n_R -1)}  \right] dt .
\end{equation}
\normalsize
To calculate (\ref{integralfinal}), the following result will be used.

Now, integral (5.6) will be calculated.
\begin{theorem} \label{teocriteriomaiorautov}
\begin{eqnarray*}
&  &  \int_{d_2}^{\infty}  \left(\frac{\Gamma(l+n_R-1, \frac{ c \cdot t -4 d_1}{4 N_0})}{\Gamma(l+n_R-1)}  \right) \\
& & \times \left[ \frac{\left( t-d_2 \right)^{n_T+n_R -2}}{ \Gamma(n_T+n_R -1)}  \exp\left(-t+d_2\right) \right] dt \\
& & = \sum_{i=0}^{l+nr-2}  \sum_{j=0}^{i} \sum_{k=0}^{i-j}  \left\{
\left(\frac{1}{4 N_0}\right)^{j+k-n_T -n_R +1} \frac{(-1)^j}{i! \cdot k!}
\left(
\begin{array}{ll}
i
\\ j
\end{array}
\right) \right.\\
& & \times (4 d_1)^{j} d_2^k  (n_T +n_R-1)_{i-j} \frac{(j-i)_k}{(j-i-n_T-n_R+2)_k} c^{i-j} \\
& & \left.\times \left(4 N_0 +c\right)^{j-i +k-n_T -n_R +1} \exp\left(\frac{- c \cdot d_
 2 +4 d_1}{4 N_0}\right) \right\},
\end{eqnarray*}
where $(x)_k$ represents the Pochhammer symbol.
\end{theorem}

Theorem \ref{teocriteriomaiorautov} presents an approximation for the error probability of the sent codeword $C$ wrongly decoded by $E$, in a transmission on a MIMO channel with a quasi-static coherent flat Rayleigh fading, where the maximum likelihood decoder is endowed with spectral norm. Thus, to obtain STC with small error probability, we need to find codes which minimize the expression in theorem \ref{teocriteriomaiorautov}. Figure \ref{minimizar} shows the graph of this expression as a function of
\[
c= \parallel  E-C \parallel_2^2,
\]
where $\parallel E-C \parallel_{2}^2 \hspace{1mm} =  \hspace{1mm} \lambda_{1} ((E-C) (E-C)^{*})$. Therefore, $\parallel  E-C \parallel_2^2 \hspace{1mm}$ must be as large as possible.


In short

\begin{theorem} \label{criterioauto}\textbf{(Largest Eigenvalue Criterion)}

To design space-time codes in MIMO communication channels with flat quasi-static coherent Rayleigh fading, we need to determine a finite family of matrices $\mathfrak{F} \subset M(n,{\bf C})$ such that
\[
min_{C_1, C_2 , C_1 \neq C_2} \{ \parallel  C_1-C_2 \parallel_2^2  \} ,
\]
is so large as possible over all codewords $C_1, C_2 \in \mathfrak{F}$.
\end{theorem}

\section{Examples}

We give some examples of STC codes.

{\bf i)} Let
\begin{eqnarray*}
\mathfrak{F} & = & \left\{
\left( \begin{array}{cc}
0 & 0 \\
0 & 1 \end{array} \right),
\left( \begin{array}{cc}
0 & 0 \\
1 & 0 \end{array} \right),
\left( \begin{array}{cc}
0 & 0 \\
1 & 1 \end{array} \right), \left( \begin{array}{cc}
0 & 1 \\
0 & 0 \end{array} \right), \right. \\
&  & \left( \begin{array}{cc}
0 & 1 \\
0 & 1 \end{array} \right),
\left( \begin{array}{cc}
1 & 0 \\
0 & 0 \end{array} \right),
\left( \begin{array}{cc}
1 & 0 \\
1 & 0 \end{array} \right),
\left( \begin{array}{cc}
1 & 1 \\
0 & 0 \end{array} \right), \\
&  & \left. \left( \begin{array}{cc}
1 & 1 \\
1 & 1 \end{array} \right) \right\}
\end{eqnarray*}
the family of all not null singular binary matrices. Using the Rank and Determinant Criterion, this set may not be used as STC, but $\|C_1 - C_2\|_2 = \sqrt{2}$.

{\bf ii)} Let $M(x,z) = \left( \begin{array}{cc}
x i & -\overline{z} \\
z & -x i \end{array} \right)$, where $x \in {\bf R}$ and $z \in {\bf C}$. All $M(x,z)$ are traceless matrices, then a finite family of these matrices may not be used as STC from Trace Criterion, but $\|M(x,z)\|_2 = \sqrt{x^2 + |z|^2}$, then for convenient choices of $x$ and $z$ we may consider finite families of these matrices where $\|M(x_1,z_1) - M(x_2,z_2)\|_2^2$ will be as large as we want.

{\bf iii)} The work \cite{Hug} studies space-time group codes. An important example is the following. Let $\mathfrak{G} = \{\pm\left( \begin{array}{cc}
1 & 0 \\
0 & 1\end{array} \right),\pm\left(\begin{array}{cc}
i & 0 \\
0 & -i\end{array} \right),\pm\left(\begin{array}{cc}
0 & 1 \\
-1 & 0\end{array} \right),\pm\left(\begin{array}{cc}
0 & i \\
i & 0\end{array} \right)\}$ and $D = \left( \begin{array}{cc}
1 & -1 \\
1 & 1\end{array} \right)$. Then $\mathfrak{C} = D\mathfrak{G}$ is called the quaternion STC code, and satisfies the Rank and Determinant Criterion. Now, if $D_1 = \left( \begin{array}{cc}
1 & 1 \\
1 & 1\end{array} \right)$ then $\mathfrak{C_1} = D_1\mathfrak{G}$ is a new STC with $\|C_1 = C_2\|_2 = 4\sqrt{2}$ for all $C_1 \neq C_2 \in \mathfrak{C_1}$, but the Rank and Determinant Criterion is not satisfied.

\section{Conclusion}
It is proposed a natural environment where the space-time codes live in and it is obtained a new design criterion of space-time codes for multi-antenna communication systems on coherent Rayleigh fading channels . The objective of this criterion is minimize the pairwise error probability of the maximum likelihood decoder, endowed with the matrix spectral norm. The random matrix theory is used, and a very useful approximation function for the probability density function of the largest eigenvalue of a Wishart Matrix, is obtained. New classes of space-time codes, which are not possible to consider by the other criteria, are given.


%

\appendices
\section{Proof of Theorem 5.6}
\begin{IEEEproof}
Since
\small
\begin{equation} \label{gamaincomplesomato}
\frac{\Gamma(l+n_R-1, \frac{c t -4 d_1}{4 N_0})}{\Gamma(l+n_R-1)} = \sum_{i=0}^{l+nr-2} \frac{\left( \frac{ c \cdot t -4 d_1}{4 N_0} \right)^i}{i!}  \exp\left(- \frac{ c t - 4 d_1}{4 N_0}\right),
\end{equation}
\normalsize
one has
\small
\begin{eqnarray*}
\lefteqn{\int_{d_2}^{\infty}  \left( \frac{\Gamma(l+n_R-1, \frac{ c \cdot t -4 d_1}{4 N_0})}{\Gamma(l+n_R-1)}  \right) \left[ \frac{\left( t-d_2 \right)^{n_T+n_R -2}}{ \Gamma(n_T+n_R -1)}  \exp\left(-t+d_2\right) \right] dt} & & \\
& = & \int_{d_2}^{\infty}  \left( \exp\left( \frac{ -c \cdot t + 4 d_1}{4 N_0}\right) \sum_{i=0}^{l+nr-2} \frac{\left( \frac{ c \cdot t -4 d_1}{4 N_0} \right)^i}{i!} \right) \\
& & \times \left[ \frac{\left( t-d_2 \right)^{n_T+n_R -2}}{ \Gamma(n_T+n_R -1)}  \exp\left(-t+d_2\right) \right] dt \\
& = & \sum_{i=0}^{l+nr-2} \frac{\exp\left( \frac{d_1}{N_0} +d_2\right)}{\Gamma(n_T+n_R -1)} \left( \frac{1}{4 N_0} \right)^i \frac{1}{i!} \int_{d_2}^{\infty}  \exp\left(- \frac{ c \cdot t}{4 N_0}\right) \\
& & \times \left(c t -4 d_1 \right)^i  \left( t-d_2 \right)^{n_T+n_R -2} \exp\left(-t \right) dt.
\end{eqnarray*}
\normalsize
From Proposition \ref{propintegr},
\small
\begin{eqnarray*}
\lefteqn{\int_{d_2}^{\infty}  \left( \frac{\Gamma(l+n_R-1, \frac{ c \cdot t -4 d_1}{4 N_0})}{\Gamma(l+n_R-1)}  \right) \left[ \frac{(t - d_2)^{n_T+n_R -2}}{ \Gamma(n_T+n_R -1)}  \exp(-t+d_2) \right] dt} & & \\
& = & \sum_{i=0}^{l+nr-2}  \left\{ \left( \frac{\exp(\frac{d_1}{N_0} + d_2)}{\Gamma(n_T+n_R -1)} \left( \frac{1}{4 N_0} \right)^i \frac{1}{i!} \right)
\sum_{j=0}^{i}
\left((-1)^j \left( \begin{array}{c}
i \\
j
\end{array}
\right) \right. \right. \\
& & c^{i-j} (4 d_1)^{j} \left(\frac{4 N_0 +c}{4 N_0}\right)^{j-i-n_T -n_R +1} \\
&  & \times \Gamma(-j+i+n_T +n_R-1) \exp\left(-\frac{d_2}{4 N_0} (c+ 4N_0)\right)  \\
&  &  \left. \times 1F1\left[ j-i, j-i-n_T -n_R+1, \frac{d_2}{4 N_0} (c+4 N_0)\right]
\right\} \\
& = &  \exp\left(\frac{- c \cdot d_2 +4 d_1}{4 N_0}\right)
\sum_{i=0}^{l+nr-2}  \left\{  \left( \frac{1}{4 N_0} \right)^i \frac{1}{i!}
\sum_{j=0}^{i}
(-1)^j
\left(
\begin{array}{c}
i
\\ j
\end {array}
\right) \right. \\
& & \times c^{i-j} (4 d_1)^{j} \left(\frac{4 N_0 +c}{4 N_0}\right)^{j-i-n_T -n_R +1}  \\
&  & \times  \frac{\Gamma(-j+i+n_T +n_R-1)}{\Gamma(n_T +n_R-1)}  \\
& & \left. \times 1F1 \left[ j-i, j-i-n_T -n_R+2, \frac{d_2}{4 N_0} (c+4 N_0)\right]
 \right\} \\
& = & \displaystyle \exp\left(\frac{- c d_2 +4 d_1}{4 N_0}\right)
\sum_{i=0}^{l+nr-2}  \left\{  \left( \frac{1}{4 N_0} \right)^i \frac{1}{i!}
\sum_{j=0}^{i}
(-1)^j
\left(
\begin{array}{c}
i
\\ j
\end {array}
\right) \right. \\
& & \times c^{i-j}   (4 d_1)^{j} \left(\frac{4 N_0 +c}{4 N_0}\right)^{j-i-n_T -n_R +1} (n_T +n_R-1)_{i-j} \\
& & \left. \times 1F1 \left[ j-i, j-i-n_T -n_R+2, \frac{d_2}{4 N_0} (c+4 N_0)\right]
 \right\} \\
& = & \exp\left(\frac{- c \cdot d_2 +4 d_1}{4 N_0}\right)
\sum_{i=0}^{l+nr-2}  \left\{  \left( \frac{1}{4 N_0} \right)^i \frac{1}{i!}
\sum_{j=0}^{i}
(-1)^j
\left(
\begin{array}{c}
i
\\ j
\end {array}
\right) \right. \\
& & \times c^{i-j} (4 d_1)^{j} \left(\frac{4 N_0 +c}{4 N_0}\right)^{j-i-n_T -n_R +1} (n_T +n_R-1)_{i-j} \\
& & \left. \times \left( \sum_{k=0}^{i-j} \frac{(j-i)_k}{(j-i-n_T-n_R +2)_k} \left(\frac{d_2}{4 N_0} (c+ 4N_0) \right)^k \frac{1}{k!}
\right)  \right\} \\
& = & \exp\left(\frac{- c \cdot d_2 +4 d_1}{4 N_0}\right)
\sum_{i=0}^{l+nr-2}  \sum_{j=0}^{i} \sum_{k=0}^{i-j}  \left\{  \left( \frac{1}{4 N_0} \right)^i \frac{1}{i!}
(-1)^j \right. \\
& & \times \left(
\begin{array}{c}
i
\\ j
\end {array}
\right)
c^{i-j}   (4 \cdot d_1)^{j} \left(\frac{4 N_0 +c}{4 N_0}\right)^{j-i-n_T -n_R +1} \\
&  & \times  (n_T +n_R-1)_{i-j}
\left( \frac{(j-i)_k}{(j-i-n_T-n_R +2)_k} \right. \\
& & \left. \left. \times \left(\frac{d_2}{4 N_0} (c+ 4N_0) \right)^k \frac{1}{k!} \right)  \right\} \\
& = & \sum_{i=0}^{l+nr-2}  \sum_{j=0}^{i} \sum_{k=0}^{i-j}  \left\{
\left(\frac{1}{4 N_0}\right)^{j+k-n_T -n_R +1} \frac{(-1)^j}{i! \cdot k!}
\left(
\begin{array}{c}
i
\\ j
\end {array}
\right)  \right. \\
& & \times (4 d_1)^{j} d_2^k  (n_T +n_R-1)_{i-j} \frac{(j-i)_k}{(j-i-n_T-n_R+2)_k} \\
& & \left. \times c^{i-j}  \left(4 N_0 +c\right)^{j-i +k-n_T -n_R +1}   \exp\left(\frac{- c \cdot d_2 +4 d_1}{4 N_0}\right)  \right\}\,,
\end{eqnarray*}
\normalsize
and the result follows.
\end{IEEEproof}

Appendix one text goes here.

\section{}

\begin{proposition}\label{propintegr}
\small
\begin{eqnarray*}
\lefteqn{\int_{d_2}^{\infty}  \exp\left(-\frac{c t}{4 N_0}\right)
(c t - 4 d_1)^i  (t - d_2)^{n_T+n_R - 2} \exp(-t ) dt} & & \\
& & = \sum_{j=0}^{i} ( (-1)^j
\left(
\begin{array}{ll}
i
\\ j
\end{array}
\right) c^{i-j} (4 d_1)^{j} \left(\frac{4 N_0 +c}{4 N_0}\right)^{j-i-n_T -n_R +1} \\
&  & \times \Gamma(-j+i+n_T +n_R-1) \exp\left(-\frac{d_2}{4 N_0} (c+ 4N_0)\right) \\
&  & \times 1F1(j - i, j-i-n_T -n_R+2, \frac{d_2}{4 N_0} (c + 4 N_0)) ) ,
\end{eqnarray*}
\normalsize
where $1F1(a,b;z)$ is the confluent hypergeometric function.
\end{proposition}

\begin{IEEEproof}
From Newton binomial formula, one has
$(ct-4d_1)^i = \sum_{j=0}^{i} (-1)^{j} \left( \begin{array}{c}
i \\
j
\end{array} \right) (c t)^{i-j} (4d_1)^j.$
Then, we need to calculate
\begin{eqnarray*}
& & \sum_{j=0}^{i} (-1)^{j} \left( \begin{array}{c}
i \\
j
\end{array} \right) c^{i-j} (4d_1)^j \int_{d_2}^{\infty}  \exp\left(- \frac{ t (c + 4 N_0)}{4 N_0}\right) \\
& & \times (t)^{i-j} \left( t-d_2 \right)^{n_T+n_R -2} dt .
\end{eqnarray*}
Since
\begin{eqnarray*}
&  &  \int_{d_2}^{\infty}  \exp\left(- \frac{ t (c + 4 N_0)}{4 N_0}\right)  (t)^{i-j}
\left( t-d_2 \right)^{n_T+n_R -2} dt \\
& = & \left( \frac{4 N_0 +c}{4 N_0} \right)^{1-i+j-n_R -n_T} \Gamma (-1+i-j+n_R +n_T) \\
& & \times \exp\left(- \frac{ d_2 (c + 4 N_0)}{4 N_0}\right) \\
&  & \times 1F1 \left[ -i+j,2-i+j-n_R -n_T, \frac{ d_2 (c + 4 N_0)}{4 N_0} \right] ,
\end{eqnarray*}
the result follows.
\end{IEEEproof}

\section*{Acknowledgment}

The authors would like to thank...

\ifCLASSOPTIONcaptionsoff
  \newpage
\fi



%

%

\begin{IEEEbiographynophoto}{Carlos A. R. Martins}
Biography text here.
\end{IEEEbiographynophoto}

\begin{IEEEbiographynophoto}{Eduardo Brandani da Silva}
Biography text here.
\end{IEEEbiographynophoto}





\end{document}